\documentclass[conference]{IEEEtran}
\usepackage[utf8]{inputenc}
\IEEEoverridecommandlockouts
\usepackage{cite}
\usepackage{amsmath,amssymb,amsfonts}
\usepackage[table,xcdraw]{xcolor}
\usepackage{algorithmic}
\usepackage{graphicx}
\usepackage{textcomp}
\usepackage{comment}
\usepackage{xcolor}
\usepackage{amsmath,amssymb,amsfonts}
\usepackage{graphicx}
\usepackage{algorithmic}
\usepackage{graphicx}
\usepackage{textcomp}
\usepackage{lipsum}
\usepackage{amsmath}
\usepackage[numbers]{natbib}
\usepackage{url}
\usepackage[section]{placeins}
\usepackage{authblk}
\usepackage{float}
\usepackage{array}
\usepackage{tabularx}
\usepackage{multirow}
\usepackage{caption}
\usepackage{multirow}
\usepackage{adjustbox}
\def\BibTeX{{\rm B\kern-.05em{\sc i\kern-.025em b}\kern-.08em
    T\kern-.1667em\lower.7ex\hbox{E}\kern-.125emX}}
  
\begin{document} 
\title{BlockTheFall: Wearable Device-based Fall Detection Framework Powered by Machine Learning and Blockchain for Elderly Care \vspace{-4mm}
}

\author[1*]{Bilash Saha}
\author[1]{Md Saiful Islam}
\author[1]{Abm Kamrul Riad}
\author[1]{Sharaban Tahora}
\author[1]{Hossain Shahriar} 
\author[2]{Sweta Sneha \vspace{-4mm} }
\affil[1]{Department of Information Technology, Kennesaw State University, Georgia, United States}
\affil[2]{Department of Information Systems and Security, Kennesaw State University, Georgia, United States}



\affil[*]{{bsaha@students.kennesaw.edu} \vspace{-5mm}}

\vspace{-6mm}

\maketitle 

\begin{abstract}
Falls among the elderly are a major health concern, frequently resulting in serious injuries and a reduced quality of life. In this paper, we propose "BlockTheFall," a wearable device-based fall detection framework which detects falls in real time by using sensor data from wearable devices. To accurately identify patterns and detect falls, the collected sensor data is analyzed using machine learning algorithms. To ensure data integrity and security, the framework stores and verifies fall event data using blockchain technology. The proposed framework aims to provide an efficient and dependable solution for fall detection with improved emergency response, and elderly individuals' overall well-being. Further experiments and evaluations are being carried out to validate the effectiveness and feasibility of the proposed framework, which has shown promising results in distinguishing genuine falls from simulated falls. By providing timely and accurate fall detection and response, this framework has the potential to substantially boost the quality of elderly care. 
\end{abstract}


\section{Introduction}

Falls have grown to be a significant global public health issue in recent years. According to WHO \cite{who}, With an estimated 684K fatalities per year, falls are the second biggest cause of unintentional injury deaths worldwide, with over 80\% of these deaths occurring in low and middle-income countries. The US Centers for Disease Control and Prevention (CDC) reports that falls among older individuals is the main reason for both fatal and nonfatal injuries. In the US alone, falls resulted in more than 39,000 fatalities and 3 million visits to emergency rooms by older people in 2019 \cite{stat1}. Additionally, Medicare and Medicaid paid for around 75\% of the direct medical expenditures associated with falls among older persons in the US in 2015 \cite{stat2}, which topped \$50 billion. Furthermore, according to the WHO, falls cause 37.3 million serious injuries that need medical attention annually. With the aging of the world's population comes a large increase in fatal falls among older persons over 60. According to Pew \cite{pew}, the senior population in the US is predicted to rise from 13.1\% in 2010 to 21.4\% in 2050. The number of people in danger of falling will dramatically rise in the future years as the aging population grows.

Falls can create post-fall anxiety syndrome, which can harm older persons' mental health in addition to causing physical damage \cite{damage}. Moreover, falls can make older persons lose faith in their ability to walk safely, which can result in self-imposed activity limitations and a loss in both physical and mental health \cite{rubenstein2006falls}. In light of this, there is an increasing demand for technologies like wearable technology and blockchain that can aid in lowering falls among older persons. The market for wearable technology has been expanding quickly in recent years and is anticipated to do so at a substantial rate going forward. The market for wearable technology was valued at \$32.63 billion in 2019 \cite{stat3} and is projected to increase at a compound annual growth rate (CAGR) of 12.6\% from 2019 to 2025, reaching \$74.03 billion (2020-2025). Accelerometers are often in combination with gyroscopes \cite{sheng2022development} with some more complicated sensors including barometric sensors and an algorithm frequently used in wearable technology for fall detection \cite{montesinos2018wearable}. These sensors can be worn on many body parts, such as the wrist, forearm, pelvis, neck, sternum, chest, thigh, cruris, shank, knee, and ankle \cite{rucco2018type}, or they can be inserted into shoes. The expanding popularity of wearable technology across a variety of sectors and the rising need for health monitoring tools like wearable fall detectors are the main drivers of this expansion.

Blockchain technology is a compelling choice for creating a safe fall detection system for aged care since it can enhance healthcare systems' security, transparency, and efficiency. The size of the worldwide blockchain technology market was estimated at \$3.67 billion in 2020 \cite{stat4} and is anticipated to grow at a CAGR of 67.3\% to reach \$39.7 billion by 2025. (2020-2025) \cite{stat5}.
In this study, we present a sensor-based low-cost fall detection method that boosts security by utilizing wearables with Android-enabled capabilities and blockchain technology. Our study's goal is to develop a trustworthy fall detection system for geriatric care. Our research's goals are:
\begin{itemize}
    \item Develop a fall detection system that can accurately distinguish between human actions including sitting, standing, and walking.
    \item Secure data transfer between wearables and family members or caregivers using blockchain technology, guaranteeing that sensitive information like fall alerts are transmitted securely and only accessible by authorized individuals ensuring HIPAA compliance \cite{HIPAA}.
    \item Compare the precision of our suggested method with other well-known supervised and unsupervised machine learning algorithms in order to assess the efficacy of the proposed mechanism.
\end{itemize}

\begin{table*}[h]
\footnotesize
\centering
\setlength{\tabcolsep}{0.5em}
{\renewcommand{\arraystretch}{1.2}

  \begin{tabularx}{\textwidth}{>{\hsize=.15\hsize\centering}X >{\hsize=.30\hsize\centering}X |>{\hsize=.17\hsize}X|>{\hsize=.17\hsize}X|>{\hsize=.25\hsize}X}

\hline
\rowcolor[HTML]{B0E0E6} 
\multicolumn{2}{|>{\hsize=.36\hsize\centering}X|}{\cellcolor[HTML]{B0E0E6}\textbf{Parameters}} &
  \centering{\begin{minipage}{1.8cm} \vspace{5pt} \textbf{Existing Fall Detection Frameworks} \vspace{5pt} \end{minipage}} &
  \centering{\begin{minipage}{1.8cm} \vspace{5pt} \textbf{Existing Wearable Fall Detection Frameworks} \vspace{5pt} \end{minipage}} &
  \begin{tabular}[c]{@{}l@{}}\textbf{Proposed Wearable }\\
  \textbf{Fall Detection Framework }\\
  \textbf{Powered By Blockchain}\end{tabular} \\ \hline
\rowcolor[HTML]{87CEEB} 
\multicolumn{1}{|l|}{\cellcolor[HTML]{87CEEB}} &
  \vspace{1pt} Cutting-edge Machine Learning Technologies &
  \vspace{1pt} Limited and narrow &
  \vspace{1pt} Limited &
  \vspace{1pt} Comprehensive and holistic \\ \cline{2-5} 
\rowcolor[HTML]{87CEEB} 
\multicolumn{1}{|l|}{\cellcolor[HTML]{87CEEB}} &
  Efficient \& Multi-perspective Progress Analysis &
  Not available &
  Not available &
  Emphasized \\ \cline{2-5} 
\rowcolor[HTML]{87CEEB} 
\multicolumn{1}{|l|}{\multirow{-3}{*}{\cellcolor[HTML]{87CEEB}
 \rotatebox{90} {
  \begin{minipage}{.9cm}
 \textbf{Analytical} \\ \textbf{power}
 \end{minipage}
 }
}} &
  Motivation to Achieve Goal &
  Limited &
  Attractive &
  Attractive \\ \hline
\rowcolor[HTML]{B0E0E6} 
\multicolumn{1}{|l|}{\cellcolor[HTML]{B0E0E6}} &
  Device Type &
  Have to carry &
  Wearable &
  Wearable \\ \cline{2-5} 
\rowcolor[HTML]{B0E0E6} 
\multicolumn{1}{|l|}{\cellcolor[HTML]{B0E0E6}} &
  Sensor used &
  Accelerometer, Gyro &
  Accelerometer, Gyro &
  Accelerometer, Gyro and Magnetometer \\ \cline{2-5} 
\rowcolor[HTML]{B0E0E6} 
\multicolumn{1}{|l|}{\cellcolor[HTML]{B0E0E6}} &
  Storage used &
  SQLite, MySQL or any relational DB &
  SQLite, MySQL or any relational DB &
  Secure Blockchain Network \\ \cline{2-5} 
\rowcolor[HTML]{B0E0E6} 
\multicolumn{1}{|l|}{\cellcolor[HTML]{B0E0E6}} &
  Leveraging Technology &
  Limited and narrow &
  Limited &
  Available \\ \cline{2-5} 
\rowcolor[HTML]{B0E0E6} 
\multicolumn{1}{|l|}{\multirow{-7}{*}{\cellcolor[HTML]{B0E0E6}

\rotatebox{90}{
  \begin{minipage}{1.8cm}
 \textbf{Solution Framework}
 \end{minipage}
}

}} &
  Access to Resources &
  Very limited &
  Available &
  Available \\ \hline
\rowcolor[HTML]{87CEEB} 
\multicolumn{1}{|l|}{\cellcolor[HTML]{87CEEB}} &
   \vspace{1pt} Patient Engagement &
   \vspace{1pt} Very limited &
   \vspace{1pt} Passive &
   \vspace{1pt} Active \\ \cline{2-5} 
\rowcolor[HTML]{87CEEB} 
\multicolumn{1}{|l|}{\cellcolor[HTML]{87CEEB}} &
  Long-term Support &
  No &
  Available &
  Available \\ \cline{2-5} 
\rowcolor[HTML]{87CEEB} 
\multicolumn{1}{|l|}{\cellcolor[HTML]{87CEEB}} &
  Access to Healthcare Services &
  Very limited &
  Limited &
  Available \\ \cline{2-5} 
\rowcolor[HTML]{87CEEB} 
\multicolumn{1}{|l|}{\multirow{-4}{*}{\cellcolor[HTML]{87CEEB}
\rotatebox{90}{
  \begin{minipage}{1.4cm}
 \textbf{Empowerment}
 \end{minipage}
}
}} &
  Relevant Health Information and Connections
  &
  Limited &
  Available &
  Comprehensive \\ \hline
\end{tabularx}%
}
\caption{Comparison of Available Fall Detection Paradigms}
\label{tab:my-table}
\end{table*}

Our proposed approach uses three primary parts to precisely detect falls and notify pre-configured contacts. The initial step in the process is collecting Accelerometer, Gyro, and Magnetometer sensor data from wearables for further analysis. The second step entails discovering how the observed fall behavior and the data were gathered. The third element includes identifying falls and notifying emergency personnel via notification services. We want to increase the security of our suggested mechanism and offer a dependable fall detection system for elderly care by utilizing blockchain technology.

In a nutshell, falls are a significant global public health issue that has an impact on the aged population. The number of people in danger of falling will dramatically rise in the future years as the aging population grows. Wearable fall detection technology has become a successful method of reducing falls and improving the quality of life for the elderly. In order to develop an accurate and secure fall detection system for elderly care, our suggested method for fall detection makes use of wearable Android-enabled devices and blockchain technology.

\section{Literature Review And Motivation}
BlockTheFall is a suggested solution that intends to solve the research gaps and issues related to healthy living in the healthcare ecosystem  shown in Table \ref{tab:my-table}, specifically for the aging population. The solution uses wearable Android-enabled smartphones and blockchain technology to deliver an accurate and secure fall detection system. The risk of falling increases with age, and these accidents can have major effects on both physical and mental health. By offering a dependable fall detection system that may alert pre-configured contacts in the event of a fall. The proposed framework tackles this problem and improves the likelihood that the user will receive prompt medical assistance. This tackles the issue of older individuals access to healthcare who may have reduced mobility or may live alone and lack convenient access to medical assistance. The major concern of the mHealth system is a public trust, and causes many data breaches and security vulnerabilities that would allow medical data alteration, unauthorised sharing, data theft, data loss, etc. Blockchain ensures the security of sensitive data storage through decentralisation, immutability, cryptography, and consensus. Due to the fact that data is stored across a distributed network of nodes, cannot be changed without network consensus, uses advanced cryptographic algorithms, and requires network approval to add new data, these features make it extremely secure \cite{qian2018towards}. Blockchain can offer a safe and impenetrable platform for exchanging health data, guaranteeing the protection of sensitive health data and enhancing the precision and speed of medical record-keeping. As the BlockTheFall framework records patient data, analyses and stores it, we integrate the blockchain network to ensure integrity of these sensitive patient data. 

BlockTheFall, A unique solution that differentiates itself from existing fall detection systems \cite{islam2020mobile} because of its incorporation of blockchain technology to secure critical health information and make it impenetrable. Individuals may easily utilize the product and seek medical aid in the case of a fall thanks to the usage of wearable Android-enabled smartphones. It differs from previous fall detection systems in that it can reliably identify falls and promptly inform pre-configured contacts, speeding up response times and improving the likelihood that the user will receive urgent medical assistance. Moreover, by addressing the difficulties and knowledge gaps in the healthcare ecosystem, its emphasis on the aging population and their particular requirements emphasizes its distinctiveness. Our system's capacity to quickly identify falls and alert emergency personnel can speed up medical response times and improve the likelihood that older persons will receive prompt care, eventually leading to better health outcomes. Also, the way older persons get treatment might be improved because of BlockTheFall's usage of wearable technology and blockchain.
BlockTheFall can improve the quality of life for older people by reducing their risk of falling and expanding their access to treatment. BlockTheFall can help create a more effective and efficient healthcare system that will benefit patients, healthcare providers, and society at large by addressing the issues and knowledge gaps in the ecosystem of healthcare linked to falls and access to treatment for older persons.

\section{APPROACH}

\subsection{Framework}
\begin{figure}[h]
\centerline{\includegraphics[width=.5\textwidth]{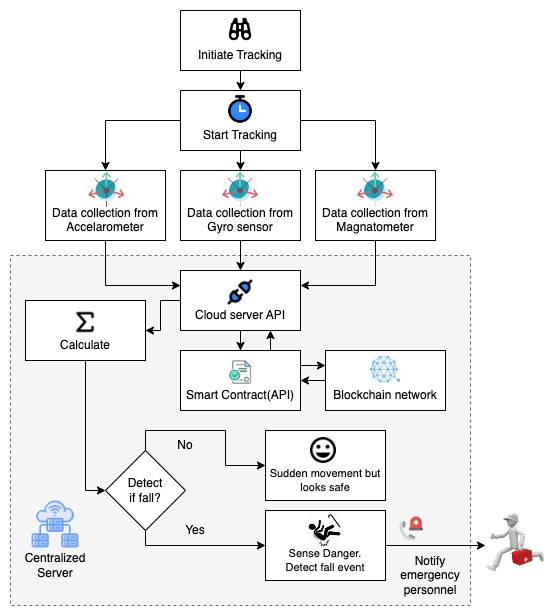}}
\caption{Secure Fall Detection BlockTheFall Framework }
\label{fig:framework}
\end{figure}

Wearable technology and machine learning algorithms are used to produce fall detection systems that have the potential to detect fall detection with precision, lower related injuries, and lower healthcare costs. As shown in Figure \ref{fig:framework}, the four key phases of the fall detection architecture are initiation, tracking, data processing, and alert generation. The user starts the fall detection application on the wearable device in the first stage, known as initiation. The program launches and waits for any sensor signals in the background.
This action is necessary to guarantee that the fall detection system is constantly operational and prepared to identify a fall. Tracking is the second stage of the framework. The wearable device's sensors, including the accelerometer, gyroscope, and magnetometer, are used by the fall detection application to follow the user's movements. The sensors track the user's motions and offer information on the device's acceleration and direction. All the sensors data determine if the user has fallen.Data processing is the third phase in the framework. A cloud server API is used to transmit the data gathered by the sensors to a central server. In order to correctly identify falls, the server analyses the data using machine learning techniques. The data processing stage is crucial because it guarantees the fall detection system's dependability and its ability to distinguish between a fall and other activities that could have motion patterns comparable to falls, including sitting or resting in bed.
Supervised learning and unsupervised learning are two machine learning methods that are used to detect falls. In supervised learning, labeled data are used to train the machine learning algorithm.
As a result, the algorithm learns to identify the characteristics of a fall based on the data that has already been categorized as a fall or not a fall. The machine learning algorithm is taught with unlabeled data in unsupervised learning. As a result, the data is not labeled, and the algorithm develops an innate understanding of the patterns in the data. The framework's fourth stage is utilizing smart contract API to store the sensor data in a safe blockchain network.The sensor data is safely and inaccessibly stored in this stage.
The immutability, transparency, and security of the data are all ensured by blockchain technology.
This step is crucial because it ensures the fall detection system is reliable. Alert generation is the framework's last phase. When a fall is discovered, the server notifies the emergency services of the accident and the user's location. The notice can be delivered through push notification, SMS, or email.
The stage of alert production is crucial since it guarantees that the user will receive quick medical care. The fall detection framework has several advantages. Early fall detection can result in prompt intervention and protect against major injuries. The ability of the system to distinguish between a fall and other activities with potentially similar motion patterns depends on accurate fall detection.
Using blockchain technology for secure data storage guarantees that the data is unchangeable, transparent, and safe. The provision of quick medical care guarantees that emergency services are informed of the fall and the user's location as soon as possible.

\subsection{Methodology}

\begin{figure*}[!h]
\centerline{\includegraphics[width=1.0\textwidth]{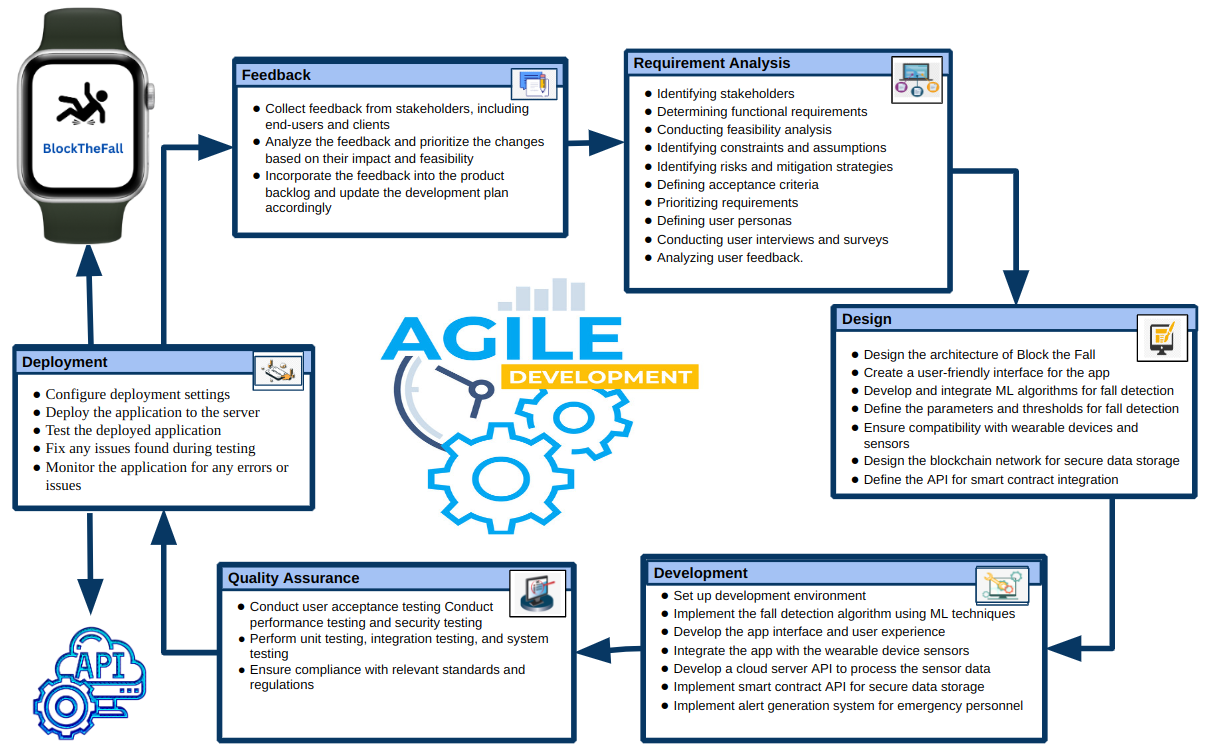}}
\caption{ Agile methodology for the development of BlockTheFall Framework }
\label{fig:agile}
\end{figure*}

\begin{figure*}[!h]
\centerline{\includegraphics[width=0.7\textwidth]{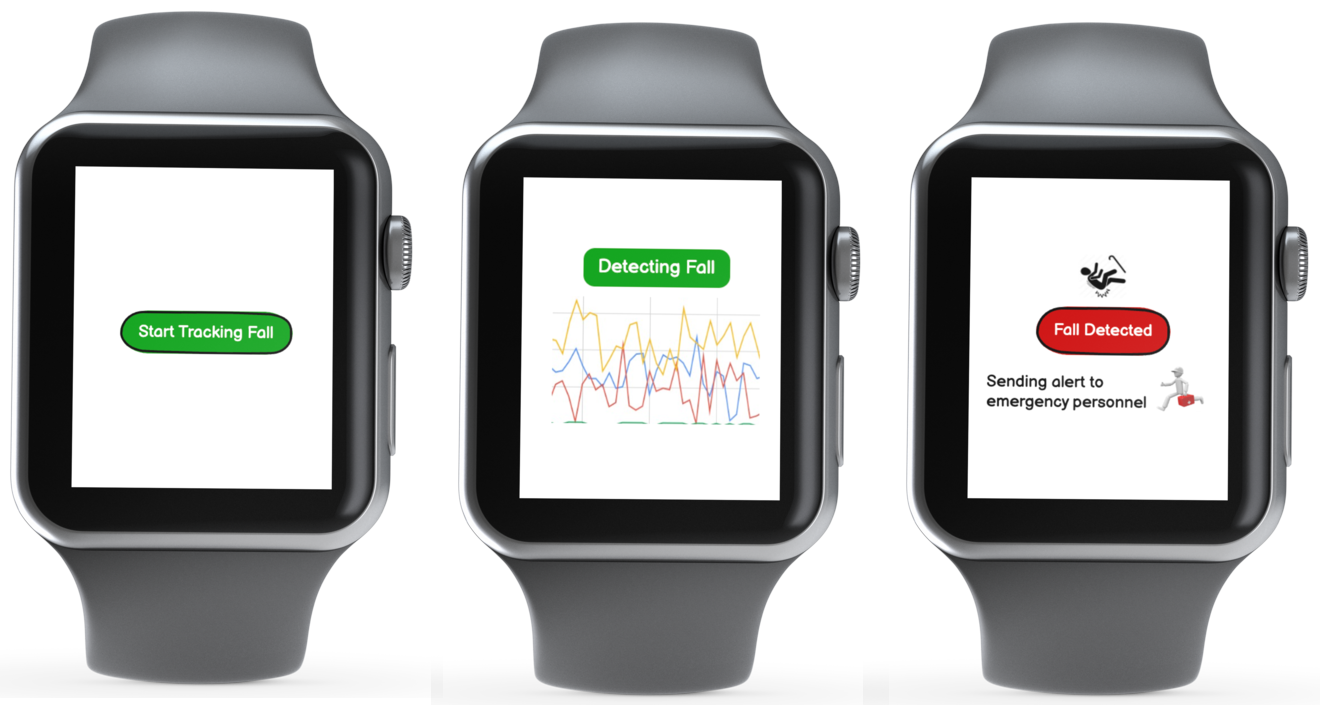}}
\caption{ Developed Watch Application in Action  }
\label{fig:application}
\end{figure*}

We begin with the requirement analysis step as we design the BlockTheFall framework utilizing Agile methodology as shown in figure \ref{fig:agile}. To identify the important aspects of the fall detection app, such as the sensors to be utilized, the types of warnings to be created, and the data processing methods to be applied, we collect and evaluate the user requirements. We proceed to the design phase after doing the requirement analysis. Here, we develop the user interface for the wearable device app as well as the general architecture of the fall detection system. We also produce a thorough development plan, complete with timeframes and objectives. The fall detection system's implementation marks the start of the development phase. The development process is divided into tiny, achievable tasks or sprints using the Agile technique. We work on the tasks in brief, iterative cycles, prioritizing them according to their relevance and complexity. We routinely test the app as it is being developed to make sure it fulfills user needs and performs as planned. Moreover, we employ automated testing methods to find and address any potential faults or problems. Following the development step, we proceed to the quality assurance phase. Here, we thoroughly evaluate the fall detection system to make sure it is dependable, precise, and secure. To make sure that the app satisfies the highest quality requirements, we employ a number of testing techniques, including functional testing, performance testing, and security testing. We launch the fall detection app on the wearable and server after the quality assurance process. We make sure the app is simple to use, accessible, and capable of smoothly integrating with the other fall detection system components. Lastly, we solicit input from users and stakeholders to ascertain whether the fall detection app fulfills their requirements and expectations. We use this input to build the next versions of the app and to make any required corrections or enhancements to the current version. 
We are able to construct the BlockTheFall application, as shown in Figure \ref{fig:application} in an adaptable, iterative, and group-based way using Agile methodology.

\section{ Results and Discussion }

\begin{figure}[h]
\centerline{\includegraphics[width=0.45\textwidth]{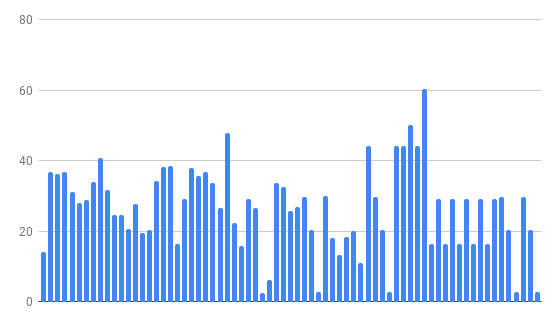}}
\caption{  Graphical Representation of Accelerometer sensor Data }
\label{fig:accelerometer}
\end{figure}

\begin{figure}[h]
\centerline{\includegraphics[width=0.45\textwidth]{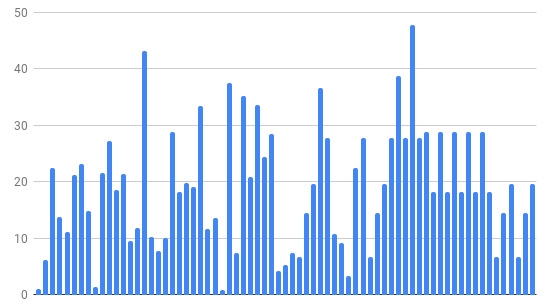}}
\caption{ Graphical Representation of Gyro Sensor Data }
\label{fig:gyro}
\end{figure}

\begin{figure}[!htbp]
\centerline{\includegraphics[width=0.45\textwidth]{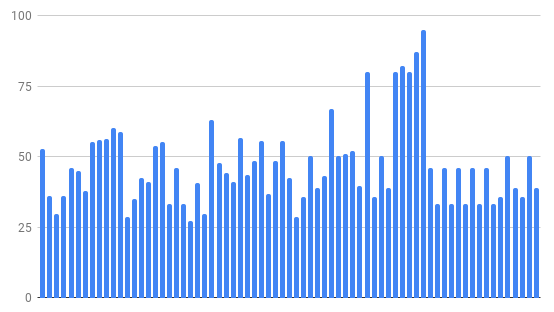}}
\caption{ Graphical Representation of Magnetormeter Sensor Data }
\label{fig:magnetormeter}
\end{figure}

\begin{figure}[!htbp]
\centerline{\includegraphics[width=0.45\textwidth]{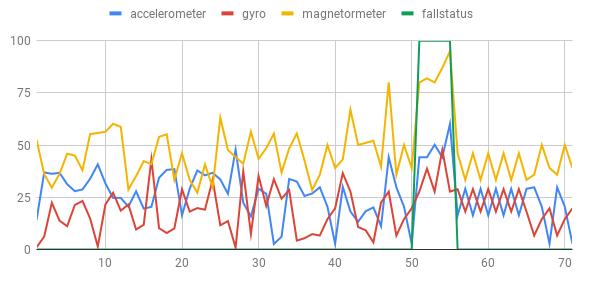}}
\caption{ Graphical Representation of Combined Sensor Data  }
\label{fig:combined}
\end{figure}

We developed a dataset with approximately 11,935 events to distinguish genuine falls from simulated falls by analyzing sensor data variations. The dataset included 3,528 actual and simulated fall events and 8,407 fake fall events artificially generated in a controlled laboratory environment. The dataset was split into subsets of 70\%/30\%, 75\%/25\%, and 80\%/20\% for training and testing purposes. The gyro sensor readings as shown in Figure \ref{fig:gyro} for actual falls ranged from 27.726 to 47.726, with an average of 33.926. The maximum gyro sensor measurement for fake falls was 43.196, with an average of 17.930. The accelerometer sensor data as illustrated in Figure \ref{fig:accelerometer}for actual falls ranged from 44.19 to 60.31, with an average of 48.59, while the range for simulated falls was from 2.60 to 47.80, with an average of 24.94. The magnetometer sensor readings as depicted in Figure \ref{fig:magnetormeter} for actual falls ranged from 80.00 to 95.00, with an average of 84.80, while the simulated falls had a range of 27.19 to 80.00, with an average of 44.371. K-means clustering was used as an unsupervised learning method to group data points into predetermined clusters based on their similarity. However, the study found that it also included some false positives in the identified clusters. The magnetometer sensor readings for actual falls ranged from 80.00 to 95.00, with an average of 84.80, while the simulated falls had a range of 27.19 to 80.00, with an average of 44.371. K-means clustering was used as an unsupervised learning method to group data points into predetermined clusters based on their similarity. Unfortunately, the investigation discovered that some of the detected clusters also had some false positives. The Naive Bayes, Linear Regression, and Neural Network supervised learning methods were used in the research to train and evaluate a dataset for the identification of falls. Figure \ref{fig:combined} shows the combined sensor data and spike detecting a real fall. 

The approaches are contrasted in Table \ref{tab:confusion} using the same dataset, with 70\% training datasets and 30\% test datasets for each method. Although logistic regression mistakenly recognized 114 fall events as not falls and 84 not falls as falls, naive bayes and neural networks incorrectly identified 48 and 66 fall events and 24 and 72 not fall events, respectively. Table \ref{tab:confusion} displays the results of using three machine learning methods on 75\% of the training dataset and 25\% of the testing dataset.  However, logistic regression misclassified 189 fall events as fake falls and 147 false falls as actual falls. Naive Bayes misidentified 112 false falls as real falls and 35 true falls as fake falls. The neural network also generated 161 inaccurate predictions of real falls as fake falls and 112 wrong predictions of fake falls as true falls. Table \ref{tab:confusion} displays the results of applying three machine learning techniques to the dataset for fall detection, which consists of 80\% training datasets and 20\% test datasets. Naive Bayes and Neural Network accurately identified 1240 and 1215 cases of fall events, and 4405 and 4355 examples, respectively, as fake fall events. The balanced accuracy results for the fall detection dataset are shown in Table \ref{tab:comparison}, which shows that K means clustering, an unsupervised learning technique, performs better than any of the three supervised learning algorithms  Naive Bayes, Logistic Regression, and Neural Networks.

\begin{table*}[!htbp]
\centering
\footnotesize
\resizebox{\textwidth}{!}{%
\begin{tabular}{|cccccccccccc|}
\hline
\rowcolor[HTML]{CBCEFB} 
\multicolumn{9}{|c|}{\cellcolor[HTML]{CBCEFB}\textbf{Supervised Machine Learning}} &
  \multicolumn{3}{c|}{\cellcolor[HTML]{CBCEFB}\textbf{Unsupervised Machine Learning}} \\ \hline
\rowcolor[HTML]{CBCEFB} 
\multicolumn{3}{|c|}{\cellcolor[HTML]{CBCEFB}\textbf{Logistic Regression}} &
  \multicolumn{3}{c|}{\cellcolor[HTML]{CBCEFB}\textbf{Naive Bayes}} &
  \multicolumn{3}{c|}{\cellcolor[HTML]{CBCEFB}\textbf{Neural Network}} &
  \multicolumn{3}{c|}{\cellcolor[HTML]{CBCEFB}\textbf{K Means Clustering}} \\ \hline
\rowcolor[HTML]{CBCEFB} 
\multicolumn{1}{|c|}{\cellcolor[HTML]{CBCEFB}Predicted: No} &
  \multicolumn{1}{c|}{\cellcolor[HTML]{CBCEFB}Predicted: Yes} &
  \multicolumn{1}{c|}{\cellcolor[HTML]{CBCEFB}N= 8328} &
  \multicolumn{1}{c|}{\cellcolor[HTML]{CBCEFB}Predicted: No} &
  \multicolumn{1}{c|}{\cellcolor[HTML]{CBCEFB}Predicted: Yes} &
  \multicolumn{1}{c|}{\cellcolor[HTML]{CBCEFB}N= 8328} &
  \multicolumn{1}{c|}{\cellcolor[HTML]{CBCEFB}Predicted: No} &
  \multicolumn{1}{c|}{\cellcolor[HTML]{CBCEFB}Predicted: Yes} &
  \multicolumn{1}{c|}{\cellcolor[HTML]{CBCEFB}N= 8328} &
  \multicolumn{1}{c|}{\cellcolor[HTML]{CBCEFB}Predicted: No} &
  \multicolumn{1}{c|}{\cellcolor[HTML]{CBCEFB}Predicted: Yes} &
  N= 8328 \\ \hline
\rowcolor[HTML]{CBCEFB} 
\multicolumn{1}{|c|}{\cellcolor[HTML]{CBCEFB}6204} &
  \multicolumn{1}{c|}{\cellcolor[HTML]{CBCEFB}114} &
  \multicolumn{1}{c|}{\cellcolor[HTML]{CBCEFB}Actual:No} &
  \multicolumn{1}{c|}{\cellcolor[HTML]{CBCEFB}6354} &
  \multicolumn{1}{c|}{\cellcolor[HTML]{CBCEFB}48} &
  \multicolumn{1}{c|}{\cellcolor[HTML]{CBCEFB}Actual:No} &
  \multicolumn{1}{c|}{\cellcolor[HTML]{CBCEFB}6318} &
  \multicolumn{1}{c|}{\cellcolor[HTML]{CBCEFB}66} &
  \multicolumn{1}{c|}{\cellcolor[HTML]{CBCEFB}Actual:No} &
  \multicolumn{1}{c|}{\cellcolor[HTML]{CBCEFB}6702} &
  \multicolumn{1}{c|}{\cellcolor[HTML]{CBCEFB}16} &
  Actual:No \\ \hline
\rowcolor[HTML]{CBCEFB} 
\multicolumn{1}{|c|}{\cellcolor[HTML]{CBCEFB}84} &
  \multicolumn{1}{c|}{\cellcolor[HTML]{CBCEFB}1926} &
  \multicolumn{1}{c|}{\cellcolor[HTML]{CBCEFB}Actual:Yes} &
  \multicolumn{1}{c|}{\cellcolor[HTML]{CBCEFB}24} &
  \multicolumn{1}{c|}{\cellcolor[HTML]{CBCEFB}1902} &
  \multicolumn{1}{c|}{\cellcolor[HTML]{CBCEFB}Actual:Yes} &
  \multicolumn{1}{c|}{\cellcolor[HTML]{CBCEFB}72} &
  \multicolumn{1}{c|}{\cellcolor[HTML]{CBCEFB}1866} &
  \multicolumn{1}{c|}{\cellcolor[HTML]{CBCEFB}Actual:Yes} &
  \multicolumn{1}{c|}{\cellcolor[HTML]{CBCEFB}21} &
  \multicolumn{1}{c|}{\cellcolor[HTML]{CBCEFB}1589} &
  Actual:Yes \\ \hline
\rowcolor[HTML]{CBCEFB} 
\multicolumn{1}{|c|}{\cellcolor[HTML]{CBCEFB}} &
  \multicolumn{1}{c|}{\cellcolor[HTML]{CBCEFB}} &
  \multicolumn{1}{c|}{\cellcolor[HTML]{CBCEFB}} &
  \multicolumn{1}{c|}{\cellcolor[HTML]{CBCEFB}} &
  \multicolumn{1}{c|}{\cellcolor[HTML]{CBCEFB}} &
  \multicolumn{1}{c|}{\cellcolor[HTML]{CBCEFB}} &
  \multicolumn{1}{c|}{\cellcolor[HTML]{CBCEFB}} &
  \multicolumn{1}{c|}{\cellcolor[HTML]{CBCEFB}} &
  \multicolumn{1}{c|}{\cellcolor[HTML]{CBCEFB}} &
  \multicolumn{1}{c|}{\cellcolor[HTML]{CBCEFB}} &
  \multicolumn{1}{c|}{\cellcolor[HTML]{CBCEFB}} &
   \\ \hline
\rowcolor[HTML]{CBCEFB} 
\multicolumn{1}{|c|}{\cellcolor[HTML]{CBCEFB}} &
  \multicolumn{10}{c|}{\cellcolor[HTML]{CBCEFB}\textbf{Confusion matrix of fall event dataset (70\%training, 30\% testing)}} &
   \\ \hline
\multicolumn{12}{|c|}{} \\ \hline
\rowcolor[HTML]{38FFF8} 
\multicolumn{9}{|c|}{\cellcolor[HTML]{38FFF8}\textbf{Supervised Machine Learning}} &
  \multicolumn{3}{c|}{\cellcolor[HTML]{38FFF8}\textbf{Unsupervised Machine Learning}} \\ \hline
\rowcolor[HTML]{38FFF8} 
\multicolumn{3}{|c|}{\cellcolor[HTML]{38FFF8}\textbf{Logistic Regression}} &
  \multicolumn{3}{c|}{\cellcolor[HTML]{38FFF8}\textbf{Naive Bayes}} &
  \multicolumn{3}{c|}{\cellcolor[HTML]{38FFF8}\textbf{Neural Network}} &
  \multicolumn{3}{c|}{\cellcolor[HTML]{38FFF8}\textbf{K Means Clustering}} \\ \hline
\rowcolor[HTML]{38FFF8} 
\multicolumn{1}{|c|}{\cellcolor[HTML]{38FFF8}Predicted: No} &
  \multicolumn{1}{c|}{\cellcolor[HTML]{38FFF8}Predicted: Yes} &
  \multicolumn{1}{c|}{\cellcolor[HTML]{38FFF8}N= 11935} &
  \multicolumn{1}{c|}{\cellcolor[HTML]{38FFF8}Predicted: No} &
  \multicolumn{1}{c|}{\cellcolor[HTML]{38FFF8}Predicted: Yes} &
  \multicolumn{1}{c|}{\cellcolor[HTML]{38FFF8}N= 11935} &
  \multicolumn{1}{c|}{\cellcolor[HTML]{38FFF8}Predicted: No} &
  \multicolumn{1}{c|}{\cellcolor[HTML]{38FFF8}Predicted: Yes} &
  \multicolumn{1}{c|}{\cellcolor[HTML]{38FFF8}N= 11935} &
  \multicolumn{1}{c|}{\cellcolor[HTML]{38FFF8}Predicted: No} &
  \multicolumn{1}{c|}{\cellcolor[HTML]{38FFF8}Predicted: Yes} &
  N= 11935 \\ \hline
\rowcolor[HTML]{38FFF8} 
\multicolumn{1}{|c|}{\cellcolor[HTML]{38FFF8}9310} &
  \multicolumn{1}{c|}{\cellcolor[HTML]{38FFF8}189} &
  \multicolumn{1}{c|}{\cellcolor[HTML]{38FFF8}Actual:No} &
  \multicolumn{1}{c|}{\cellcolor[HTML]{38FFF8}9422} &
  \multicolumn{1}{c|}{\cellcolor[HTML]{38FFF8}112} &
  \multicolumn{1}{c|}{\cellcolor[HTML]{38FFF8}Actual:No} &
  \multicolumn{1}{c|}{\cellcolor[HTML]{38FFF8}9345} &
  \multicolumn{1}{c|}{\cellcolor[HTML]{38FFF8}161} &
  \multicolumn{1}{c|}{\cellcolor[HTML]{38FFF8}Actual:No} &
  \multicolumn{1}{c|}{\cellcolor[HTML]{38FFF8}9702} &
  \multicolumn{1}{c|}{\cellcolor[HTML]{38FFF8}89} &
  Actual:No \\ \hline
\rowcolor[HTML]{38FFF8} 
\multicolumn{1}{|c|}{\cellcolor[HTML]{38FFF8}147} &
  \multicolumn{1}{c|}{\cellcolor[HTML]{38FFF8}2289} &
  \multicolumn{1}{c|}{\cellcolor[HTML]{38FFF8}Actual:Yes} &
  \multicolumn{1}{c|}{\cellcolor[HTML]{38FFF8}35} &
  \multicolumn{1}{c|}{\cellcolor[HTML]{38FFF8}2366} &
  \multicolumn{1}{c|}{\cellcolor[HTML]{38FFF8}Actual:Yes} &
  \multicolumn{1}{c|}{\cellcolor[HTML]{38FFF8}112} &
  \multicolumn{1}{c|}{\cellcolor[HTML]{38FFF8}2317} &
  \multicolumn{1}{c|}{\cellcolor[HTML]{38FFF8}Actual:Yes} &
  \multicolumn{1}{c|}{\cellcolor[HTML]{38FFF8}51} &
  \multicolumn{1}{c|}{\cellcolor[HTML]{38FFF8}2093} &
  Actual:Yes \\ \hline
\rowcolor[HTML]{38FFF8} 
\multicolumn{1}{|c|}{\cellcolor[HTML]{38FFF8}} &
  \multicolumn{1}{c|}{\cellcolor[HTML]{38FFF8}} &
  \multicolumn{1}{c|}{\cellcolor[HTML]{38FFF8}} &
  \multicolumn{1}{c|}{\cellcolor[HTML]{38FFF8}} &
  \multicolumn{1}{c|}{\cellcolor[HTML]{38FFF8}} &
  \multicolumn{1}{c|}{\cellcolor[HTML]{38FFF8}} &
  \multicolumn{1}{c|}{\cellcolor[HTML]{38FFF8}} &
  \multicolumn{1}{c|}{\cellcolor[HTML]{38FFF8}} &
  \multicolumn{1}{c|}{\cellcolor[HTML]{38FFF8}} &
  \multicolumn{1}{c|}{\cellcolor[HTML]{38FFF8}} &
  \multicolumn{1}{c|}{\cellcolor[HTML]{38FFF8}} &
   \\ \hline
\rowcolor[HTML]{38FFF8} 
\multicolumn{1}{|c|}{\cellcolor[HTML]{38FFF8}} &
  \multicolumn{10}{c|}{\cellcolor[HTML]{38FFF8}\textbf{Confusion matrix of fall event dataset (75\%training, 25\% testing)}} &
   \\ \hline
\multicolumn{12}{|c|}{} \\ \hline
\rowcolor[HTML]{67FD9A} 
\multicolumn{9}{|c|}{\cellcolor[HTML]{67FD9A}\textbf{Supervised Machine Learning}} &
  \multicolumn{3}{c|}{\cellcolor[HTML]{67FD9A}\textbf{Unsupervised Machine Learning}} \\ \hline
\rowcolor[HTML]{67FD9A} 
\multicolumn{3}{|c|}{\cellcolor[HTML]{67FD9A}\textbf{Logistic Regression}} &
  \multicolumn{3}{c|}{\cellcolor[HTML]{67FD9A}\textbf{Naive Bayes}} &
  \multicolumn{3}{c|}{\cellcolor[HTML]{67FD9A}\textbf{Neural Network}} &
  \multicolumn{3}{c|}{\cellcolor[HTML]{67FD9A}\textbf{K Means Clustering}} \\ \hline
\rowcolor[HTML]{67FD9A} 
\multicolumn{1}{|c|}{\cellcolor[HTML]{67FD9A}Predicted: No} &
  \multicolumn{1}{c|}{\cellcolor[HTML]{67FD9A}Predicted: Yes} &
  \multicolumn{1}{c|}{\cellcolor[HTML]{67FD9A}N= 5695} &
  \multicolumn{1}{c|}{\cellcolor[HTML]{67FD9A}Predicted: No} &
  \multicolumn{1}{c|}{\cellcolor[HTML]{67FD9A}Predicted: Yes} &
  \multicolumn{1}{c|}{\cellcolor[HTML]{67FD9A}N= 5695} &
  \multicolumn{1}{c|}{\cellcolor[HTML]{67FD9A}Predicted: No} &
  \multicolumn{1}{c|}{\cellcolor[HTML]{67FD9A}Predicted: Yes} &
  \multicolumn{1}{c|}{\cellcolor[HTML]{67FD9A}N= 5695} &
  \multicolumn{1}{c|}{\cellcolor[HTML]{67FD9A}Predicted: No} &
  \multicolumn{1}{c|}{\cellcolor[HTML]{67FD9A}Predicted: Yes} &
  N= 5695 \\ \hline
\rowcolor[HTML]{67FD9A} 
\multicolumn{1}{|c|}{\cellcolor[HTML]{67FD9A}4290} &
  \multicolumn{1}{c|}{\cellcolor[HTML]{67FD9A}130} &
  \multicolumn{1}{c|}{\cellcolor[HTML]{67FD9A}Actual:No} &
  \multicolumn{1}{c|}{\cellcolor[HTML]{67FD9A}4405} &
  \multicolumn{1}{c|}{\cellcolor[HTML]{67FD9A}45} &
  \multicolumn{1}{c|}{\cellcolor[HTML]{67FD9A}Actual:No} &
  \multicolumn{1}{c|}{\cellcolor[HTML]{67FD9A}4355} &
  \multicolumn{1}{c|}{\cellcolor[HTML]{67FD9A}55} &
  \multicolumn{1}{c|}{\cellcolor[HTML]{67FD9A}Actual:No} &
  \multicolumn{1}{c|}{\cellcolor[HTML]{67FD9A}4349} &
  \multicolumn{1}{c|}{\cellcolor[HTML]{67FD9A}23} &
  Actual:No \\ \hline
\rowcolor[HTML]{67FD9A} 
\multicolumn{1}{|c|}{\cellcolor[HTML]{67FD9A}100} &
  \multicolumn{1}{c|}{\cellcolor[HTML]{67FD9A}1175} &
  \multicolumn{1}{c|}{\cellcolor[HTML]{67FD9A}Actual:Yes} &
  \multicolumn{1}{c|}{\cellcolor[HTML]{67FD9A}5} &
  \multicolumn{1}{c|}{\cellcolor[HTML]{67FD9A}1240} &
  \multicolumn{1}{c|}{\cellcolor[HTML]{67FD9A}Actual:Yes} &
  \multicolumn{1}{c|}{\cellcolor[HTML]{67FD9A}70} &
  \multicolumn{1}{c|}{\cellcolor[HTML]{67FD9A}1215} &
  \multicolumn{1}{c|}{\cellcolor[HTML]{67FD9A}Actual:Yes} &
  \multicolumn{1}{c|}{\cellcolor[HTML]{67FD9A}15} &
  \multicolumn{1}{c|}{\cellcolor[HTML]{67FD9A}1308} &
  Actual:Yes \\ \hline
\rowcolor[HTML]{67FD9A} 
\multicolumn{1}{|c|}{\cellcolor[HTML]{67FD9A}} &
  \multicolumn{1}{c|}{\cellcolor[HTML]{67FD9A}} &
  \multicolumn{1}{c|}{\cellcolor[HTML]{67FD9A}} &
  \multicolumn{1}{c|}{\cellcolor[HTML]{67FD9A}} &
  \multicolumn{1}{c|}{\cellcolor[HTML]{67FD9A}} &
  \multicolumn{1}{c|}{\cellcolor[HTML]{67FD9A}} &
  \multicolumn{1}{c|}{\cellcolor[HTML]{67FD9A}} &
  \multicolumn{1}{c|}{\cellcolor[HTML]{67FD9A}} &
  \multicolumn{1}{c|}{\cellcolor[HTML]{67FD9A}} &
  \multicolumn{1}{c|}{\cellcolor[HTML]{67FD9A}} &
  \multicolumn{1}{c|}{\cellcolor[HTML]{67FD9A}} &
   \\ \hline
\rowcolor[HTML]{67FD9A} 
\multicolumn{1}{|c|}{\cellcolor[HTML]{67FD9A}} &
  \multicolumn{10}{c|}{\cellcolor[HTML]{67FD9A}\textbf{Confusion matrix of fall event dataset (80\%training, 20\% testing)}} &
   \\ \hline
\end{tabular}%
}
\caption{Confusion matrix of fall event }
\label{tab:confusion}
\end{table*}

\begin{table}[h]
\centering
\footnotesize
\begin{tabular}{|
>{\columncolor[HTML]{C0C0C0}}c |
>{\columncolor[HTML]{9AFF99}}c |
>{\columncolor[HTML]{FFFFC7}}c |
>{\columncolor[HTML]{CBCEFB}}c |
>{\columncolor[HTML]{ECF4FF}}c |}
\hline
\textbf{\begin{tabular}[c]{@{}c@{}}Dataset \\ split\end{tabular}} &
  \textbf{\begin{tabular}[c]{@{}c@{}}Naive \\ Bayes\end{tabular}} &
  \textbf{\begin{tabular}[c]{@{}c@{}}Logistic \\ Regression\end{tabular}} &
  \textbf{\begin{tabular}[c]{@{}c@{}}Neural \\ Network\end{tabular}} &
  \textbf{\begin{tabular}[c]{@{}c@{}}K Means \\ Clustering\end{tabular}} \\ \hline
(70\%/30\%) & 0.89324178 & 0.79732965 & 0.91700292 & 0.95700292 \\ \hline
(75\%/25\%) & 0.84304717 & 0.74843305 & 0.89672206 & 0.92672206 \\ \hline
(80\%/20\%) & 0.83149622 & 0.78715737 & 0.90497069 & 0.93497069 \\ \hline
\end{tabular}%
\caption{Balanced Accuracy and comparison with Naive Bayes, Logistic Regression, Neural Network and K Means Clustering}
\label{tab:comparison}
\end{table}

\section{Conclusion}
The development of reliable and economical fall detection techniques is essential since falls among senior populations continue to be a major problem. This research suggests a revolutionary fall detection technique using wearable devices and blockchain technology ensures security and accuracy. This study shows how blockchain-enabled technology and using wearable devices have the potential to change secured aged care and raise the standard of living for seniors. Further study in this field may lead to the creation of less expensive and more effective fall detection systems that will help senior citizens receive home care.  

\section*{Acknowledgment}
\footnotesize
The work is supported by the National Science Foundation
under NSF Award \#2100115, \#2209638.
Any opinions, findings, recommendations, expressed in this
material are those of the authors and do not necessarily reflect
the views of the National Science Foundation.

\renewcommand{\bibfont}{\footnotesize}

\bibliographystyle{IEEEtranN}
\bibliography{biblography.bib}
\end{document}